\documentclass[aps,pre]{revtex4}
\usepackage{amsmath}
\usepackage{graphicx}
\usepackage{dcolumn}
\usepackage{bm}
\usepackage{epsfig}
\usepackage{graphics}%
\usepackage{amsfonts}%
\usepackage{amssymb}

\begin{document}

\title{ On dissipationless shock waves in a discrete nonlinear Schr\"odinger equation}
\author{A.M. Kamchatnov$^{1,2}$,  A. Spire$^{2}$,  V.V. Konotop$^{2}$}
\affiliation{$^{1}$Institute of Spectroscopy, Russian Academy of Sciences, 
Troitsk, Moscow
Region, 142190, Russia }
\affiliation{$^{2}$
Departamento de F\'{\i}sica and Centro de F\'{\i}sica Te\'orica e
Computacional, Universidade de Lisboa, Av.~Prof.~Gama Pinto 2, Lisbon
1649-003, Portugal}
\date{\today }

\begin{abstract}
It is shown that the generalized discrete nonlinear Schr\"odinger equation can
be reduced in a small amplitude approximation to the 
KdV, mKdV, KdV(2) or
the fifth-order KdV equations, depending on values of the parameters. In dispersionless
limit these equations lead to wave breaking phenomenon for general enough
initial conditions, and, after taking into account small dispersion effects,
result in formation of dissipationless shock waves. The Whitham theory of
modulations of nonlinear waves is used for 
analytical description of such waves.

\vspace*{0.3cm}


\end{abstract}
\maketitle


\section{Introduction}

Dissipationless shock waves have been experimentally observed or their
existence has been theoretically predicted in various nonlinear media -- water
\cite{Whitham74}, plasma \cite{Sagdeev64}, optical fibers \cite{Krokel},
lattices \cite{lattice}. In contrast to usual dissipative shocks where
combined action of nonlinear and dissipation effects leads to sharp jumps of
the wave intensity, accompanied by abrupt changes of other wave
characteristics, in dissipationless shocks the viscosity effect is negligibly
small compared with the dispersive one, and, instead of intensity jumps, the
combined action of nonlinear and dispersion effects leads to formation
oscillatory wave region (for review see e.g. \cite{kamch2000}). Since
intrinsic discreteness of a solid state system gives origin to strong
dispersion, which can dominate dissipative effects in wave phenomena, it is of
considerable interest to investigate details of formation of dissipationless
shocks in lattices.

As a model, in the present paper we choose the general discrete nonlinear
Schr\"{o}dinger (GDNLS) equation
\begin{equation}
i\dot{q}_{n}+(1-\eta|q_{n}|^{2})(q_{n+1}+q_{n-1}-2q_{n})+2(\rho^{2}%
-|q_{n}|^{2})q_{n}=0\label{eq1}%
\end{equation}
introduced by Salerno \cite{Salerno92a,ESSE92}. Eq. (\ref{eq1}) turned out to
be an important model not only because of its property to provide a
one-parametric transition between an integrable Ablowitz-Ladik (AL) model and
the so-called discrete nonlinear Schr\"{o}dinger (DNLS) equation ($\eta=1$ and
$\eta=0$, respectively), but also because of a number of physical application
for a review of which we refer to \cite{Hennig}.

Obviously, (\ref{eq1}) has a constant amplitude solution $q_{n}=\rho$. It was
shown in \cite{KS97a,KS97b}, that in a small amplitude, $|a|\ll\rho$, and long
wave (so that the discrete site index $n$ can be replaced by a continuous
coordinate $x$) limit, evolution of small amplitude perturbations with respect
to this constant background,
\begin{equation}
q_{n}(t)=(\rho+a(x,t))\exp(-i\phi(x,t)),\label{eq2}%
\end{equation}
is governed by the Korteweg-de Vries (KdV) equation for the amplitude
$a(x,t)$:
\begin{equation}
a_{t}-\frac{2(3-4\eta\rho^{2})}{\sqrt{1-\eta\rho^{2}}}aa_{x}+\frac{\sqrt
{1-\eta\rho^{2}}}{12\rho}[3(1-\eta\rho^{2})-\rho^{2}]a_{xxx}=0\label{eq3}%
\end{equation}
which is written in the reference system moving with velocity $2\rho
\sqrt{1-\eta\rho^{2}}$ of linear waves in dispersionless limit. It is well
known (see, e.g. \cite{Whitham74}) that if the initial pulse is strong enough,
so that the nonlinear term dominates over the dispersive one at the initial
stage of the pulse evolution, then the dissipationless shock wave develops
after the wave breaking point. The theory of such waves, described by the KdV
equation, is well developed (see, e.g. \cite{kamch2000}). Existence of the
respective shock waves for model (\ref{eq1}) has been predicted analytically
and observed in numerical simulations in \cite{KS97a,KS97b}. Moreover, as it
is shown in \cite{KonChaos} a nonlinear Schr\"{o}dinger equation of a rather
general type can bear ``KdV-type'' shock waves. In this context the results
presented below in the present paper although being mostly related to model
(\ref{eq1}) display some general characteristic features of a lattice of a
nonlinear Schr\"{o}dinger type.

The coefficients of (\ref{eq3}) depend on two parameters $\eta$ and $\rho$ and
can vanish at special choice of these parameters, so the KdV equation looses
its applicability for these values of $\eta$ and $\rho$. Physically acceptable
values of $\eta$ and $\rho$ are limited by the inequalities
\begin{equation}
\label{eq4}0\leq\eta\leq\mathrm{min}\{1,1/\rho^{2}\},\quad0<\rho<\infty,
\end{equation}
and this region of the $(\rho,\eta)$ plane is depicted in Fig.~1.

\begin{figure}[h]
\label{fig0} \centerline{
\rotatebox{270}{\scalebox{.50}%
[0.50]{\includegraphics{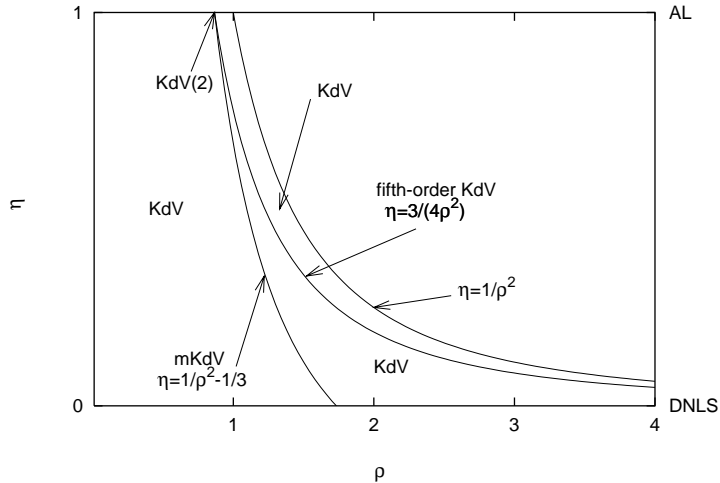}}}
} \caption{Diagram showing different
continuous limits of the GDNLS equation. The abbreviations are explained in
the text.}%
\end{figure}

Along the line
\begin{equation}
\label{eq5}\eta=\frac{3}{4\rho^{2}}%
\end{equation}
the coefficient before the nonlinear term in (\ref{eq3}) vanishes. This means
that implied in derivation of the KdV equation approximation limited to only
quadratic nonlinearities fails and higher nonlinearities have to be taken into
account for accurate description of the wave dynamics. Hence, one may expect
that the modified KdV (mKdV) equation with cubic nonlinearity arises for
values of $\rho$ and $\eta$ related by (\ref{eq5}).

Along the line
\begin{equation}
\label{eq6}\eta=\frac{1}{\rho^{2}}-\frac{1}{3}%
\end{equation}
the coefficient before the dispersion term in (\ref{eq3}) vanishes what means
that higher order dispersion effects have to be taken into account. In this
case one may expect that evolution equation for $a(x,t)$ contains quadratic
nonlinear term and linear dispersion term with fifth-order space derivative of
$a(x,t)$.

The most interesting point corresponds to the values
\begin{equation}
\label{eq7}\eta=1,\quad\rho=\sqrt{3}/2
\end{equation}
when both nonlinear and dispersion coefficients vanish. Since Eq.~(\ref{eq1})
with $\eta=1$ coincides with the completely integrable Ablowitz-Ladik
equation, one may expect that in a small amplitude approximation this equation
with the parameters equal to (\ref{eq7}) reduces again to a completely
integrable equation. The KdV equation (\ref{eq3}) with $\eta=1$ is valid for
all interval $0<\rho<1$ except for some vicinity of the point $\rho=\sqrt
{3}/2$, and therefore one can suppose that at the point (\ref{eq3}) one has to
obtain the second equation of the KdV hierarchy KdV(2) in which the higher
order nonlinear and dispersion effects play the dominant role.

The aim of this paper is two-fold. At first, in Sec.~\ref{small_amplitude} we
derive the evolution equations for the whole region (\ref{eq4}) and show that
along the line (\ref{eq5}) the small-amplitude approximation reduces to the
mKdV equation (Sec~\ref{mKdV}), along the line (\ref{eq6}) to 
a nonlinear equation with the dispersion of the 
fifth-order (Sec~\ref{Kawa}), and at the point (\ref{eq7}) to the KdV(2) equation
(Sec~\ref{KdV2}). The last result sheds some new light on the nature of higher
equations of the KdV hierarchy---they arise as small amplitude approximations
to completely integrable equations, if lower orders of nonlinear and
dispersion contributions vanish at some values of the parameters of the
equation under consideration.

The second aim of the paper is to develop a theory of shock waves
(Sec~\ref{shock}) for the mKdV (Sec~\ref{shock_mKdV})and KdV(2)
(Sec~\ref{shock_KdV2}) equations analogous to that developed earlier for the
KdV equation. This theory permits one to described in details the behavior of
shocks after the wave breaking point for different values of the parameters
entering into Eq.~(\ref{eq1}). The results are summarized in Conclusion.

\section{Small amplitude approximation}

\label{small_amplitude}

Using ansatz (\ref{eq2}) and replacing a discrete index $n$ by a continuous
variable $x$, we rewrite Eq.~(\ref{eq1}) as
\begin{equation}
\label{eq8}%
\begin{split}
ia_{t}+(\rho+a)\phi_{t}-4\rho^{2}a-6\rho a^{2} +[1-\eta(\rho+a)^{2}]\\
\times\{(\rho+a(x+1,t))\exp[-i(\phi(x+1,t)-\phi(x,t))]\\
+(\rho+a(x-1,t))\exp[-i(\phi(x-1,t)-\phi(x,t))]\}=0.
\end{split}
\end{equation}
In the linear approximation, this equation yields for the harmonic wave
solution
\[
a(x,t)\propto\exp[i(Kx-\Omega t)], \quad\phi(x,t)\propto\exp[i(Kx-\Omega t)],
\]
the dispersion relation \cite{KS97a,KS97b}
\begin{equation}
\label{eq9}%
\begin{split}
\Omega=\pm4\sqrt{1-\eta\rho^{2}}\sin\frac{K}{2}\left[ \rho^{2}+(1-\eta\rho
^{2}) \sin^{2}\frac{K}{2}\right] ^{1/2}\\
\cong\pm2\rho\sqrt{1-\eta\rho^{2}}\,K\cdot\left[ 1+ \frac{3(1-\eta\rho
^{2})-\rho^{2}}{24\rho^{2}}\,K^{2}+O(K^{4})\right] ,
\end{split}
\end{equation}
where expansion in powers of $K$ corresponds to taking into account different
orders of the dispersion effects. In the lowest order, when the dispersion
effects are neglected, linear waves propagate with constant velocity
\begin{equation}
\label{eq10}v=\pm2\rho\sqrt{1-\eta\rho^{2}}.
\end{equation}
To evaluate contribution of small (for $|a|\ll\rho$) nonlinear effects, it is
convenient to introduce a small parameter $\varepsilon\sim a$ and pass to such
scaled variables in which nonlinear and dispersion effects make contributions
of the same order of magnitude into evolution of the wave. Since the choice of
these scaled variables depends on values of the parameters, we shall consider
the relevant cases separately.

\subsection{KdV equation}

\label{KdV}

For the sake of completeness we start by reproducing briefly some results of
\cite{KS97a}. We expand $a(x\pm1,t)$ and $\phi(x\pm1,t)$ into the Taylor
series around $x$, introduce scaling indexes
\begin{equation}
a\sim\varepsilon,\quad t\sim\varepsilon^{-\alpha},\quad x\sim\varepsilon
^{-\beta},\quad\phi\sim\varepsilon^{\gamma},\label{eq11}%
\end{equation}
and demand that in the reference frame moving with velocity (\ref{eq10}) of
linear waves the lowest quadratic nonlinearity has the same order of magnitude
as the second term in the expansion (\ref{eq9}) of the dispersion relation,
\[
a\sim\phi_{x},\quad a_{t}\sim aa_{x}\sim a_{xxx},
\]
which yield
\[
\alpha=\tfrac{3}{2},\quad\beta=\gamma=\tfrac{1}{2}.
\]
Thus, the scaled variables have the form
\begin{equation}
\tau=\varepsilon^{3/2}t,\quad\xi=\varepsilon^{1/2}(x+vt),\quad v=\pm2\rho
\sqrt{1-\eta\rho^{2}},\label{eq12}%
\end{equation}
and $a(x,t)$ and $\phi(x,t)$ should be looked for in the form of expansions
\begin{equation}%
\begin{split}
a &  =\varepsilon a^{(1)}+\varepsilon^{2}a^{(2)}+\varepsilon^{3}a^{(3)}%
+\ldots,\\
\phi &  =\varepsilon^{1/2}\phi^{(1)}+\varepsilon^{3/2}\phi^{(2)}%
+\varepsilon^{5/2}\phi^{(2)}\ldots
\end{split}
\label{eq13}%
\end{equation}
Then in the lowest order in expansion of Eq.~(\ref{eq8}) in powers of
$\varepsilon$ we obtain the relationship
\begin{equation}
\phi_{\xi}^{(1)}=\frac{4\rho}{v}a^{(1)},\label{eq14}%
\end{equation}
 where we have chosen the
upper sign of $v$ in (\ref{eq12}), and in the next order the KdV equation
(\ref{eq3}) written in terms of the scaled variables. 
The nonlinear term
changes its sign at (\ref{eq5}) and the dispersion term 
changes its sign at
(\ref{eq6}). Hence, there can be as bright solitons 
against a background ($a(x,t)>0$) or dark 
solitons ($a(x,t)<0$) of the GDNLS
equation approximated by the KdV equation (\ref{eq3}). 
In both cases it is
possible to pass to new dependent variable 
\begin{equation}
a  =-\frac{1-\eta\rho^{2}}{12\rho-16\eta\rho^{3}}
[3(1-\eta\rho^2)-\rho^2]u
\end{equation}
and change time as $t \rightarrow\frac{\sqrt{1-\eta\rho^{2}}}{12\rho}[3(1-\eta\rho^{2}
)-\rho^{2}]\ t$
such that the KdV equation takes the
standard form
\begin{equation}
u_{t}+6uu_{x}+u_{xxx}=0.\label{eq15}%
\end{equation}

\subsection{mKdV equation}

\label{mKdV}

At $\eta=3/4\rho^{2}$ the quadratic nonlinearity in Eq. (\ref{eq3}) becomes
zero and in order to describe the wave in the vicinity of the breaking point
the scaling should be chosen to take into account cubic nonlinearity which is
now must have the same order of magnitude as $a_{xxx}$,
\[
a\sim\phi_{x},\quad a_{t}\sim a^{2}a_{x}\sim a_{xxx}.
\]
Then, using scaling (\ref{eq11}) we find
\[
\alpha=3,\quad\beta=1,\quad\gamma=0,
\]
so that instead of (\ref{eq12}) we have the following scaled variables
\begin{equation}
\tau=\varepsilon^{3}t,\quad\xi=\varepsilon(x+vt),\quad v=\pm\rho,\label{eq16}%
\end{equation}
where value of velocity is found by substitution of (\ref{eq5}) into
(\ref{eq10}), and instead of the second expansion in (\ref{eq13}) we have
\[
\phi=\phi^{(1)}+\varepsilon\phi^{(2)}+\varepsilon^{2}\phi^{(3)}+\ldots.
\]
Then in the lowest order in expansion of Eq.~(\ref{eq8}) in powers of
$\varepsilon$ we obtain
\begin{equation}
\phi_{\xi}^{(1)}=\frac4v a^{(1)}\label{eq17}%
\end{equation}
what coincides with Eq.~(\ref{eq14}) after substitution of $v=\rho$. In the
next order we get the relationship
\begin{equation}
\phi_{\xi}^{(2)}=4a^{(2)}+\frac{6}{\rho}a^{(1)2},\label{eq18}%
\end{equation}
and finally in the highest relevant order we obtain the mKdV equation
\begin{equation}
a_{\tau}^{(1)}+\left(  4+\frac{21}{\rho^{2}}\right)  \rho\,a^{(1)2}a_{\xi
}^{(1)}+\frac{1}{32\rho}\left(  1-\frac{4\rho^{2}}{3}\right)  a_{\xi\xi\xi
}^{(1)}=0.\label{eq19}%
\end{equation}
Since along the line (\ref{eq5}) we have $\rho>\sqrt{3}/2$, the coefficient
before the dispersion term is always negative and hence Eq.~(\ref{eq19}) can
be transformed to the following standard form
\begin{equation}
u_{t}-6u^{2}u_{x}+u_{xxx}=0,\label{eq20}%
\end{equation}
where
\begin{equation}
a^{(1)}   =\frac14\sqrt{\frac{(4\rho^{2}-3)}
{(4\rho^2+21)}}u
\end{equation}
and we renormalize the time variable
\begin{equation}
t \rightarrow\frac{1}{32\rho}(1-4\frac{\rho^{2}}{3})t\ .
\end{equation}

\subsection{Fifth-order KdV equation}

\label{Kawa}

At $\eta=1/\rho^{2}-1/3$ the first order dispersion effects in Eq. (\ref{eq3})
disappears and in this case scaling should be chosen so that the quadratic
nonlinearity has the order of magnitude of $a^{(V)}$,
\[
a\sim\phi_{x},\quad a_{t}\sim aa_{x}\sim a^{(V)},
\]
which yields
\[
\alpha=\tfrac54,\quad\beta=\tfrac14,\quad\gamma=\tfrac34,
\]
so that the scaled variables are given by
\begin{equation}
\label{eq21}\tau=\varepsilon^{5/4}t,\quad\xi=\varepsilon^{1/4}(x+vt),\quad
v=\pm\frac{2\rho^{2}}{\sqrt{3}},
\end{equation}
where velocity $v$ is found by substitution of (\ref{eq6}) into (\ref{eq10}),
and now the condition of cancellation of terms in the second order demands
that expansions of $a$ and $\phi$ have the form
\begin{equation}
\label{eq21a}%
\begin{split}
a=\varepsilon a^{(1)}+\varepsilon^{3/2}a^{(2)}+\varepsilon^{2}a^{(3)}%
+\ldots,\\
\phi=\varepsilon^{3/4} \phi^{(1)}+\varepsilon^{5/4}\phi^{(2)}+\varepsilon
^{7/4}\phi^{(2)}\ldots
\end{split}
\end{equation}
Then in the lowest order in expansion of Eq.~(\ref{eq8}) in powers of
$\varepsilon^{1/2}$ we get again
\begin{equation}
\label{eq22}\phi^{(1)}_{\xi}=\frac{2\sqrt{3}}{\rho}a^{(1)}%
\end{equation}
what can be obtained from Eq.~(\ref{eq14}) by substitution of (\ref{eq6}). In
the next order we get the relationship
\begin{equation}
\label{eq23}\phi^{(2)}_{\xi}=\frac{2\sqrt{3}}{\rho}a^{(2)}- \frac{1}{2\sqrt
{3}\rho}a^{(1)}_{\xi\xi}%
\end{equation}
and finally in the highest relevant order we obtain the equation
\begin{equation}
\label{eq24}a^{(1)}_{\tau}+\frac{2\sqrt{3}}{\rho}\left( 1-\frac{4\rho^{2}}%
{3}\right) a^{(1)}a^{(1)}_{\xi} +\frac{\sqrt{3}\rho^{2}}{270}a^{(1)}_{\xi
\xi\xi\xi\xi}=0
\end{equation}
Here the nonlinear term can be
obtained from the corresponding term in the KdV equation (\ref{eq3}) by
substitution of (\ref{eq6}) and dispersion term reproduces the expansion of
the dispersion relation (\ref{eq9}) at the same value of $\eta$.

\subsection{KdV(2) equation}

\label{KdV2}

At the point (\ref{eq7}) both the quadratic nonlinear and first order
dispersion terms disappear, so that now the cubic nonlinearity must be of the
same order of magnitude as $a^{(V)}$,
\[
a\sim\phi_{x},\quad a_{t}\sim a^{2}a_{x}\sim a^{(V)},
\]
which yield
\[
\alpha=\tfrac52,\quad\beta=\gamma=\tfrac12,
\]
and the scaled variables are given by
\begin{equation}
\label{eq25}\tau=\varepsilon^{5/2}t,\quad\xi=\varepsilon^{1/2}(x+vt),\quad
v=\pm\frac{\sqrt{3}}{2},
\end{equation}
where $v$ is the velocity of linear waves at the point (\ref{eq7}). The
variables $a(x,t)$ and $\phi(x,t)$ have the same form of expansions
(\ref{eq13}) as in the KdV equation case. In the lowest order we obtain, as
one should expect, Eq.~(\ref{eq17}); in the next order we get the
relationship
\begin{equation}
\label{eq26}\phi^{(2)}_{\xi}=4a^{(2)}+4\sqrt{3}\,a^{(1)2}-\tfrac13a^{(2)}%
_{\xi\xi};
\end{equation}
and at the last relevant order we obtain the equation
\begin{equation}
\label{eq27}a^{(1)}_{\tau}+16\sqrt{3}a^{(1)2}a^{(1)}_{\xi}-\tfrac
43\,a^{(1)}_{\xi} a^{(1)}_{\xi\xi}-\tfrac23\,a^{(1)}a^{(1)}_{\xi\xi\xi}
+\frac{\sqrt{3}}{360}a^{(1)}_{\xi\xi\xi\xi\xi}=0.
\end{equation}
As one should expect, the main nonlinear term here coincides with that of the
mKdV equation (\ref{eq19}) at the point (\ref{eq7}), and linear dispersion
term with the corresponding term of the fifth-order KdV equation (\ref{eq24}) at the
same point.

By means of replacements
\[
a^{(1)}=-\frac{1}{8\sqrt{3}}u,\quad\tau=-30\sqrt{3}\,t,\quad\xi=x
\]
Eq.~(\ref{eq27}) can be transformed to standard form of the second equation of
the KdV hierarchy---KdV(2) (see, e.g. \cite{kamch2000}):
\begin{equation}
\label{eq28}u_{t}=\tfrac{15}2u^{2}u_{x}+5u_{x}u_{xx}+\tfrac52uu_{xxx}%
+\tfrac14u^{(V)}%
\end{equation}

Thus, the completely integrable AL equation reduces in the small amplitude
approximation either to the KdV equation beyond some vicinity of the point
(\ref{eq7}), or to the second equation of the KdV hierarchy at the point
(\ref{eq7}), so that approximate equations remain completely integrable in
both cases. This observation suggests that the property of complete
integrability preserves in framework of singular perturbation scheme, which is
a known phenomenon (see e.g. ~\cite{ZK}). Then higher equations of some
hierarchy may arise as approximate equations. This happens if at some values
of the parameters of the underline completely integrable problem nonlinearity
and, hence, dispersion of lower equations of the hierarchy vanish. This
phenomenon can be viewed as physical meaning of the higher equations of
hierarchies of integrable equations.

\section{Dissipationless shock waves}

\label{shock}

In dispersionless limit when dispersion effects can be neglected compared with
nonlinear ones, all derived above equations reduce in the leading
approximation to the Hopf-like equation
\begin{equation}
\label{eq29}u_{t}+u^{n}u_{x}=0,
\end{equation}
where $n=1$ for the KdV and fifth-order KdV equations and $n=2$ for mKdV and KdV(2)
equations. It is well-known (see, e.g. \cite{kamch2000}) that Eq. (\ref{eq29})
with general enough initial condition leads to formation of the wave breaking
point after which the solution becomes multi-valued function of $x$. This
means that near the wave breaking point one cannot neglect the dispersion
effects. If we take them into account, then the multi-valued region is
replaced by the oscillatory region of the solution of the full equation. This
oscillatory region is called dissipationless shock wave and its analytical
description is the aim of this section.

Existing theory of dissipationless shock waves can be applied in principle to
completely integrable equations only. Among equations derived in the preceding
section, however, fifth-order KdV equation (\ref{eq24}) does not belong to this
class. Fortunately, just this case of zero first-order dispersion was studied
numerically in \cite{KS97a,KS97b}. We also bear in mind that the
dissipationless shock waves of the KdV equation are already described in
literature (see e.g. \cite{kamch2000}). Therefore we shall not consider this
equation here and concentrate our attention on the completely integrable models.

The analytical approach is based on the idea that the oscillatory region of
the dissipationless shock wave can be represented as a modulated periodic
solution of the equation under consideration. If the parameters defining the
solution change little on a distance of one wavelength and during the time of
order of one period, one can distinguish two scales of time in this problem --
fast oscillations of the wave and slow change of the parameters of the wave.
Then equations which govern a slow evolution of the parameters can be averaged
over fast oscillations what leads to the so-called Whitham equations
\cite{Whitham74} and their solution subject to appropriate initial and
boundary conditions describes evolution of the dissipationless shock wave.
This approach was suggested in \cite{GP73} and now it is well developed for
the KdV equation case (see, e.g. \cite{kamch2000}). The results of this theory
can be applied to shocks in GDNLS equation when it is reduced to the KdV
equation (\ref{eq3}) or (\ref{eq15}). Since this theory is presented in detail
in \cite{kamch2000}, we shall develop first analogous theory for the mKdV and
KdV(2) equations and after that compare the results obtained for different equations.

\subsection{Dissipationless shock wave in the mKdV equation (\ref{eq20})}

\label{shock_mKdV}

At first we have to express a periodic solution of the mKdV equation
(\ref{eq20}) in a form suitable for the Whitham modulation theory. Such a form
is provided automatically by the finite-gap integration method which is used
here to find the one-phase periodic solution of the mKdV equation.

The finite-gap integration method (see, e.g. \cite{kamch2000}) is based on the
complete integrability of the mKdV equation, following from a possibility to
represent this equation as a compatibility condition $\Psi_{xt}=\Psi_{tx}$ of
two linear systems
\begin{equation}%
\begin{array}
[c]{c}%
\Psi_{x}=\mathbb{U}\Psi,\quad\Psi_{t}=\mathbb{V}\Psi,\quad\Psi=\left(
\begin{array}
[c]{c}%
\psi_{1}\\
\psi_{2}%
\end{array}
\right)  ,\\
\mathbb{U}=\left(
\begin{array}
[c]{cc}%
-i\lambda & iu\\
-iu & i\lambda
\end{array}
\right)  ,\quad\mathbb{V}=\left(
\begin{array}
[c]{cc}%
A & B\\
C & -A
\end{array}
\right)  ,\\
A=-4i\lambda^{3}-2iu^{2}\lambda,\quad B=4iu\lambda^{2}-2u_{x}\lambda
-iu_{xx}+2iu^{3},\quad C=-4iu\lambda^{2}-2u_{x}\lambda+iu_{xx}-2iu^{3},
\end{array}
\label{eq30}%
\end{equation}
where $\lambda$ is a free spectral parameter. The linear systems (\ref{eq30})
have two basis solutions $\Psi^{\pm}=(\psi_{1}^{\pm},\psi_{2}^{\pm})$, from
which we build the so-called `squared basis functions'
\begin{equation}
f=-\frac{i}{2}(\psi_{1}^{+}\psi_{2}^{-}+\psi_{1}^{-}\psi_{2}^{+}),\quad
g=\psi_{1}^{+}\psi_{1}^{-},\quad h=-\psi_{2}^{+}\psi_{2}^{-}.\label{eq31}%
\end{equation}
They satisfy the following linear systems
\begin{equation}%
\begin{array}
[c]{l}%
f_{x}=-ug-uh,\quad g_{x}=-2uf-2i\lambda g,\quad h_{x}=-2uf+2i\lambda h,
\end{array}
\label{eq32}%
\end{equation}
and
\begin{equation}%
\begin{array}
[c]{l}%
f_{t}=-iCg+iBh,\quad g_{t}=2iBf+2Ag,\quad h_{t}=-2iCf-2Ah,
\end{array}
\label{eq33}%
\end{equation}
and have the following integral
\begin{equation}
f^{2}-gh=P(\lambda)\label{eq34}%
\end{equation}
independent of $x$ and $t$. The periodic solutions are distinguished by the
condition that $P(\lambda)$ be a polynomial in $\lambda$ and we shall see that
one-phase solution corresponds to the sixth degree polynomial in even powers
of $\lambda$,
\begin{equation}
P(\lambda)=\prod_{i=1}^{3}(\lambda^{2}-\lambda_{i}^{2})=\lambda^{6}%
-s_{1}\lambda^{4}+s_{2}\lambda^{2}-s_{3}.\label{eq35}%
\end{equation}
Then $f,\,g,\,h,$ satisfying (\ref{eq32})-(\ref{eq34}) should be also
polynomials in $\lambda$,
\begin{equation}%
\begin{array}
[c]{l}%
f=\lambda^{3}-f_{1}\lambda,\quad g=iu(\lambda-\mu_{1})(\lambda-\mu_{2}),\quad
h=-iu(\lambda+\mu_{1})(\lambda+\mu_{2}),
\end{array}
\label{eq36}%
\end{equation}
where $\mu_{j}$ are new dependent variables. Substitution of (\ref{eq36}) into
(\ref{eq34}) gives the conservation laws ($s_{j}$ are constants)
\begin{equation}%
\begin{array}
[c]{l}%
2f_{1}+u^{2}=s_{1},\quad f_{1}^{2}+u^{2}(\mu_{1}^{2}+\mu_{2}^{2})=s_{2},\quad
u^{2}\mu_{1}^{2}\mu_{2}^{2}=s_{3},
\end{array}
\label{eq37}%
\end{equation}
and substitution into (\ref{eq32}) and (\ref{eq33}) yields the following
important formulae
\begin{equation}
u_{x}=2iu(\mu_{1}+\mu_{2}),\label{eq38}%
\end{equation}%
\begin{equation}
u_{t}=2(2f_{1}+u^{2})u_{x}.\label{eq39}%
\end{equation}
>From (\ref{eq39}) and the first equation (\ref{eq37}) we see that $u$ depends
only on the phase
\begin{equation}
u=u(\theta),\quad\theta=x+2s_{1}t,\label{eq40}%
\end{equation}
and from (\ref{eq38}) and the other equations (\ref{eq37}) we find
\begin{equation}
u_{\theta}^{2}=u^{4}-2s_{1}u^{2}\mp8\sqrt{s_{3}}u+s_{1}^{2}-4s_{2}\equiv
Q(u),\label{eq41}%
\end{equation}
where the zeroes $u_{i}$ of the polynomial $Q(u)$ are related with the zeroes
$\lambda_{i}$ of the polynomial $P(\lambda)$ by the formulae
\begin{equation}%
\begin{array}
[c]{l}%
u_{1}=\pm(\lambda_{1}+\lambda_{2}+\lambda_{3}),\quad u_{2}=\pm(\lambda
_{1}-\lambda_{2}-\lambda_{3}),\quad u_{3}=\pm(-\lambda_{1}+\lambda_{2}%
-\lambda_{3}),\quad u_{4}=\pm(-\lambda_{1}-\lambda_{2}+\lambda_{3}).
\end{array}
\label{eq42}%
\end{equation}
If we order the zeroes $\lambda_{i}$ according to
\begin{equation}
\lambda_{1}>\lambda_{2}>\lambda_{3}>\lambda_{4},\label{eq43}%
\end{equation}
then for the upper choice of the sign in (\ref{eq41}) and (\ref{eq42}) we
have
\begin{equation}
u_{1}>u_{2}>u_{3}>u_{4}\label{eq44}%
\end{equation}
and $u$ oscillates within the interval
\begin{equation}
u_{3}\leq u\leq u_{2},\label{eq45}%
\end{equation}
where $Q(u)\geq0$. For the lower choice of the sign in (\ref{eq41}) and
(\ref{eq42}) we have
\begin{equation}
u_{1}<u_{2}<u_{3}<u_{4}\label{eq46}%
\end{equation}
and $u$ oscillates within the interval
\begin{equation}
u_{2}\leq u\leq u_{3}.\label{eq47}%
\end{equation}
We are interested in wave trains against positive constant background which
corresponds to the lower choice of sign in (\ref{eq41}) and (\ref{eq42}). In
this case Eq.~(\ref{eq41}) yields the periodic solution
\begin{equation}
u(\theta)=\frac{(u_{3}-u_{1})u_{2}-(u_{3}-u_{2})u_{1}{\mathrm{sn}}^{2}%
(\sqrt{(u_{4}-u_{2})(u_{3}-u_{1})}\,\theta/2,m)}{u_{3}-u_{1}-(u_{3}%
-u_{2}){\mathrm{sn}}^{2}(\sqrt{(u_{4}-u_{2})(u_{3}-u_{1})}\,\theta
/2,m)},\label{eq48}%
\end{equation}
where
\begin{equation}
m=\frac{(u_{3}-u_{2})(u_{4}-u_{1})}{(u_{4}-u_{2})(u_{3}-u_{1})}=\frac{\lambda
_{1}^{2}-\lambda_{2}^{2}}{\lambda_{1}^{2}-\lambda_{3}^{2}},\quad
\theta=x+2s_{1}t=x+2(\lambda_{1}^{2}+\lambda_{2}^{2}+\lambda_{3}%
^{2})t.\label{eq49}%
\end{equation}
At $\lambda_{2}=\lambda_{3}$, when $m=1$, the solution (\ref{eq48}) transforms
into soliton solution of the mKdV equation
\begin{equation}
u_{s}(\theta)=\lambda_{1}-\frac{2(\lambda_{1}^{2}-\lambda_{2}^{2})}%
{\lambda_{1}-\lambda_{2}+2\lambda_{2}\cosh^{2}(2\sqrt{\lambda_{1}^{2}%
-\lambda_{2}^{2}}\,\theta)},\quad\theta=x+2(\lambda_{1}^{2}+2\lambda_{2}%
^{2})t.\label{eq50}%
\end{equation}

In a modulated wave the parameters $\lambda_{i}$ become slow functions of $x$
and $t$. It is convenient to introduce new variables
\begin{equation}
\label{eq51}r_{1}=\lambda_{3}^{2},\quad r_{2}=\lambda_{2}^{2},\quad
r_{3}=\lambda_{1}^{2},
\end{equation}
so that Whitham equations can be written in the form (see, e.g.
\cite{kamch2000})
\begin{equation}
\label{eq52}\frac{\partial r_{i}}{\partial t}+v_{i}(r)\frac{\partial r_{i}%
}{\partial x}=0, \quad v_{i}=\left( 1-\frac{L}{\partial_{i}L}\partial
_{i}\right) V,\quad i=1,2,3,
\end{equation}
where $V=-2(r_{1}+r_{2}+r_{3})$ is the phase velocity of the nonlinear wave
(\ref{eq48}) and
\begin{equation}
\label{eq53}L=\frac{{\mathrm{K}} (m)}{\sqrt{r_{3}-r_{1}}},\quad m=\frac{r_{3}%
-r_{2}}{r_{3}-r_{1}}%
\end{equation}
is the wavelength.

Now our task is to consider the solution of the mKdV equation after the
wave-breaking point. As it follows from (\ref{eq19}), before this point in
dispersionless approximation the evolution of the pulse obeys the Hopf
equation
\begin{equation}
u_{t}-6u^{2}u_{x}=0\label{eq54}%
\end{equation}
with the well-known solution
\begin{equation}
x+6u^{2}t=f(u^{2}),\label{eq55}%
\end{equation}
where $f(u^{2})$ is determined by the initial condition. At the wave-breaking
point, which will be assumed to be $t=0$, the profile $u^{2}(x)$ has an
inflexion point with vertical tangent line,
\[
\left.  \frac{\partial x}{\partial u^{2}}\right|  _{t=0}=0,\quad\left.
\frac{\partial^{2}x}{\partial(u^{2})^{2}}\right|  _{t=0}=0.
\]
Hence, in its vicinity we can represent (\ref{eq55}) as
\begin{equation}
x+6u^{2}t=-(u^{2}-u_{b}^{2})^{3},\label{eq56}%
\end{equation}
where $u_{b}=u(x_{b},t_{b})$. Note that the
mKdV equation is not Galileo invariant and therefore we cannot eliminate the
constant parameter $u_{b}^{2}$, in contrast to the case of a KdV equation
(see, e.g. \cite{kamch2000}).

\begin{figure}[h]
\label{fig1} \centerline{
{\includegraphics{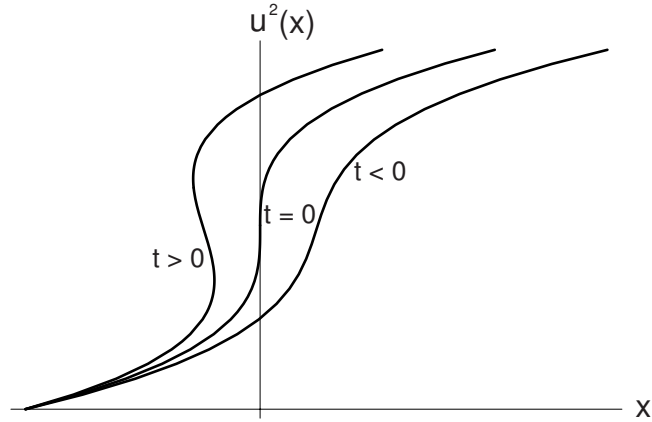}}} \caption{Formation of
multi-valued solution of the mKdV equation in dispersionless limit
(\ref{eq54}). The initial data correspond to the cubic curve $x=-(u^{2}%
-u_{b}^{2})^{3}$ (see Eq.~(\ref{eq56})). }%
\end{figure}

For $t>0$ solution (\ref{eq56}) becomes multi-valued function of $x$.
Formation of this multi-valued region is shown in Fig.~2. For $t\ge0$
we cannot neglect dispersion and have to consider full mKdV equation. Due to
effect of dispersion the multi-valued region is replaced by the region of fast
oscillations which can be represented as a modulated periodic solution of the
mKdV equation (\ref{eq20}). We rewrite this solution [see Eqs.~(\ref{eq48}),
(\ref{eq49})] in terms of the slowly varying functions $r_{i}(x,t)$,
$i=1,2,3$,
\begin{equation}
\label{eq57}u(x,t)=\frac{(\sqrt{r_{3}}+\sqrt{r_{1}})(\sqrt{r_{2}} +\sqrt
{r_{1}}-\sqrt{r_{3}})+(\sqrt{r_{3}}-\sqrt{r_{2}}) (\sqrt{r_{1}}+\sqrt{r_{2}%
}+\sqrt{r_{3}}) {\mathrm{sn}} ^{2}(2\sqrt{r_{3}-r_{1}}\,\theta,m)}{\sqrt
{r_{1}}+\sqrt{r_{3}}-(\sqrt{r_{3}}-\sqrt{r_{2}}) {\mathrm{sn}} ^{2}%
(2\sqrt{r_{3}-r_{1}}\,\theta,m)},
\end{equation}
where
\begin{equation}
\label{eq58}m=\frac{r_{3}-r_{2}}{r_{3}-r_{1}}, \quad\theta=x+2s_{1}%
t=x+2(r_{1}+r_{2}+r_{3})t.
\end{equation}
and functions $r_{i}(x,t)$ are governed by the Whitham equations (\ref{eq52}).
We have to find such solution of these equations that the region of
oscillations matches at its end points corresponding to $m=0$ and $m=1$ to the
dispersionless solution (\ref{eq56}) which we rewrite in the form
\begin{equation}
\label{eq59}x+6rt=-(r-r_{b})^{3},\quad r=u^{2},\quad r_{b}=u_{b}^{2}.
\end{equation}
This means that the solution of Eqs.~(\ref{eq52}) written in implicit form
\begin{equation}
\label{eq60}x-v_{i}(r)t=w_{i}(r),\quad i=1,2,3,
\end{equation}
must satisfy the boundary conditions
\begin{equation}
\label{eq61}\left. v_{1}\right| _{r_{2}=r_{3}}=-6r_{1},\quad\left.
v_{3}\right| _{r_{2}=r_{1}}=-6r_{3};
\end{equation}
\begin{equation}
\label{eq62}\left. w_{1}\right| _{r_{2}=r_{3}}=-(r_{1}-r_{b})^{3},\quad\left.
w_{3}\right| _{r_{2}=r_{1}}=-(r_{1}-r_{b})^{3}.
\end{equation}
Then, as we shall see from the results, mean values of $u$ will match at these
boundaries to the solution of dispersionless mKdV equation.

To find the solution (\ref{eq60}) subject to the boundary conditions
(\ref{eq61}),(\ref{eq62}), we shall follow the method developed earlier for
KdV equation (see, e.g. \cite{kamch2000}). We look for $w_{i}$ in the form
similar to (\ref{eq52}),
\begin{equation}
w_{i}=\left(  1-\frac{L}{\partial_{i}L}\partial_{i}\right)  W,\quad
i=1,2,3,\label{eq63}%
\end{equation}
and find that $W$ satisfies the Euler-Poisson equation
\begin{equation}
\partial_{ij}W-\frac{1}{2(r_{i}-r_{j})}\left(  \partial_{i}W-\partial
_{j}W\right)  =0,\quad i\neq j.\label{eq64}%
\end{equation}
For our aim it is enough to know a particular solution of this linear equation
$W=\mathrm{const}/\sqrt{P(r)}$, where $P(r)$ is a polynomial with zeroes
$r_{i}$ and it can be identified with polynomial (\ref{eq35}) with taking into
account Eqs.~(\ref{eq51}). The series expansion of this solution in inverse
powers of $r$,
\begin{equation}
W=\frac{-4r^{3/2}}{\sqrt{(r-r_{1})(r-r_{2})(r-r_{3})}}=\sum_{n=0}^{\infty
}\frac{W^{(n)}}{r^{n}},\label{eq65}%
\end{equation}
can be considered as generating function of sequence of solutions,
\begin{equation}
W^{(1)}=-2s_{1},\quad W^{(2)}=2s_{2}-\tfrac{3}{2}s_{1}^{2},\quad
W^{(3)}=3s_{1}s_{2}-2s_{3}-\tfrac{5}{4}s_{1}^{3},\label{eq66}%
\end{equation}
where $s_{1},s_{2},s_{3}$ are the coefficients of the polynomial (\ref{eq35})
expressed in terms of $r_{1},r_{2},r_{3}$:
\begin{equation}
s_{1}=r_{1}+r_{2}+r_{3},\quad s_{2}=r_{1}r_{2}+r_{1}r_{3}+r_{2}r_{3},\quad
s_{3}=r_{1}r_{2}r_{3}.\label{eq67}%
\end{equation}
It is easy to find that the resulting velocities
\begin{equation}
w_{i}^{(n)}=\left(  1-\frac{L}{\partial_{i}L}\partial_{i}\right)
W^{(n)},\quad i=1,2,3,\label{eq68}%
\end{equation}
have the following limiting values
\begin{equation}
\left.  w_{1}^{(1)}\right|  _{r_{2}=r_{3}}\equiv\left.  v_{1}\right|
_{r_{2}=r_{3}}-6r_{1},\quad\left.  w_{3}^{(1)}\right|  _{r_{2}=r_{1}}%
\equiv\left.  v_{3}\right|  _{r_{2}=r_{1}}=-6r_{3};\label{eq69}%
\end{equation}%
\begin{equation}
\left.  w_{1}^{(2)}\right|  _{r_{2}=r_{3}}=-\tfrac{15}{2}r_{1}^{2}%
,\quad\left.  w_{3}^{(2)}\right|  _{r_{2}=r_{1}}=-\tfrac{15}{2}r_{3}%
^{2};\label{eq70}%
\end{equation}%
\begin{equation}
\left.  w_{1}^{(3)}\right|  _{r_{2}=r_{3}}=-\tfrac{35}{4}r_{1}^{3}%
,\quad\left.  w_{3}^{(3)}\right|  _{r_{2}=r_{1}}=-\tfrac{35}{4}r_{3}%
^{3}.\label{eq71}%
\end{equation}
Thus, we see that if we take
\begin{equation}
w_{i}(r)=-r_{b}^{3}-\tfrac{1}{2}r_{b}^{2}w_{i}^{(1)}(r)+\tfrac{2}{5}r_{b}%
w_{i}^{(2)}(r)-\tfrac{4}{35}w_{i}^{(3)}(r),\quad i=1,2,3,\label{eq72}%
\end{equation}
then formulae (\ref{eq60}) satisfy all necessary conditions and define
dependence of $r_{1},r_{2},r_{3}$ on $x$ and $t$ in implicit form. In
Fig.~3 we have shown the dependence of $r_{1},r_{2},r_{3}$ on $x$ at
$t=0.5$ and $r_{b}=5$. It is clearly seen that $r_{2}$ and $r_{1}$ coalesce at
the right boundary $x_{+}$, where $m=1$, and $r_{2}$ and $r_{3}$ coalesce at
the left boundary $x_{-}$, where $m=0$. Dispersionless solution is depicted by
dashed line and $r_{1}$ matches this solution at $x_{-}$ and $r_{3}$ matches
it at $x_{+}$.

\begin{figure}[h]
\label{fig2} \centerline{
{\includegraphics{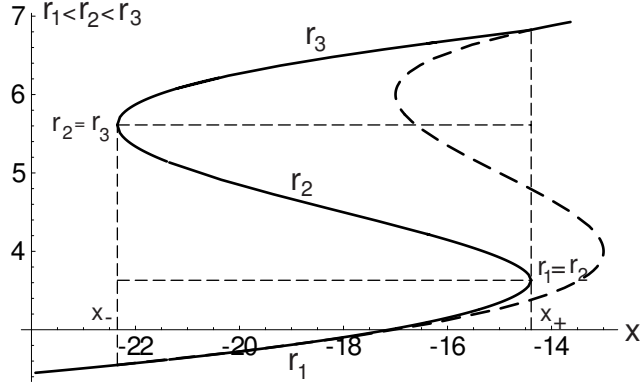}}} \caption{Dependence
of the Riemann invariants $r_{1},r_{2},r_{3}$ on $x$ at some fixed value of
time. The plots are calculated according to formulae (\ref{eq60}) with
$r_{b}=5$ and $t=0.5$. A dashed line represents the dispersionless solution
which matches the Riemann invariants at the boundaries $x_{\pm} $ of the
region of oscillations.}%
\end{figure}

Let us find the laws of motion $x_{\pm}(t)$ of the boundaries of the region of
oscillations. At the right boundary we have the condition
\begin{equation}
\label{eq73}\left. \frac{dx}{dr_{1}}\right| _{m=1}=\left. \frac{dx}{dr_{2}%
}\right| _{m=1}=0,
\end{equation}
which yields the expression
\begin{equation}
\label{eq74}t=\tfrac{12}{35}r_{1}^{2}+\tfrac4{35}r_{1}r_{3}+\tfrac3{70}%
r_{3}^{2}-\tfrac45r_{1}r_{b} -\tfrac15r_{3}r_{b}+\tfrac12r_{b}^{2},
\end{equation}
and substitution of this expression into Eqs.~(\ref{eq60}) with $r_{1}=r_{2}$
gives the coordinate $x_{+}$ expressed in terms of the Riemann invariants
$r_{1}$ and $r_{3}$:
\begin{equation}
\label{eq75}x_{+}=\tfrac1{35}\left[ -32r_{1}^{3}+2r_{3}^{3}-8r_{1}r_{3}%
(r_{3}-7r_{b})-8r_{1}^{2}(4r_{3}-7r_{b}) -7r_{3}^{2}r_{b}-35r_{b}^{3}\right] .
\end{equation}
On the other hand, this value of $x_{+}$ must coincide with coordinate
obtained from the dispersionless solution (\ref{eq59}) with $t$ equal to
Eq.~(\ref{eq74}),
\begin{equation}
\label{eq76}x_{+}=-6r_{3}t-(r_{3}-r_{b})^{3}=\tfrac1{35}\left[ -72r_{1}%
^{2}r_{3}+26r_{3}^{3}-24r_{1}r_{3} (r_{3}-7r_{b})-63r_{3}^{2}r_{b}-35r_{b}%
^{3}\right] .
\end{equation}
Comparison of these two expressions for $x_{+}$ yields the relation between
$r_{1},\,r_{2}$ and $r_{3}$ at $m=1$:
\begin{equation}
\label{eq77}\left( 4r_{1}+3r_{3}\right) _{m=1}=7r_{b}.
\end{equation}
Substitution of $r_{1}$ obtained from this equation into Eq.~(\ref{eq74})
gives
\begin{equation}
\label{eq78}t=\tfrac3{20}(r_{3}-r_{b})^{2}.
\end{equation}
Hence
\begin{equation}
\label{eq79}\left. r_{3}\right| _{m=1}=r_{b}+\tfrac23\sqrt{15t}%
\end{equation}
and again with the use of Eq.~(\ref{eq77}) we obtain
\begin{equation}
\label{eq80}\left. r_{1}\right| _{m=1}=\left. r_{2}\right| _{m=1}=r_{b}%
-\tfrac12\sqrt{15t}.
\end{equation}
These formulae give values of the Riemann invariants at the right boundary as
functions of time $t$. Their substitution into (\ref{eq75}) or (\ref{eq76})
yields the motion law of the right boundary
\begin{equation}
\label{eq81}x_{+}(t)=-6r_{b}t+\tfrac43\sqrt{\tfrac53}\,t^{3/2}.
\end{equation}

In a similar way at the left boundary $x_{-}$ the conditions
\begin{equation}
\left.  \frac{dx}{dr_{2}}\right|  _{m=0}=\left.  \frac{dx}{dr_{3}}\right|
_{m=0}=0\label{eq82}%
\end{equation}
yield
\begin{equation}
t=-\tfrac{1}{30}r_{1}^{2}-\tfrac{4}{15}r_{1}r_{3}+\tfrac{4}{5}r_{3}^{2}%
+\tfrac{1}{3}r_{1}r_{b}-\tfrac{4}{3}r_{3}r_{b}+\tfrac{1}{2}r_{b}%
^{2}\label{eq83}%
\end{equation}
which substitution into Eqs.~(\ref{eq60}) and (\ref{eq59}) gives,
respectively,
\begin{equation}
x_{-}=\tfrac{1}{5}\left[  -2r_{1}^{3}-32r_{3}^{3}+8r_{1}r_{3}(4r_{3}%
-5r_{b})+40r_{3}^{2}r_{b}-r_{b}^{3}+r_{1}^{2}(-8r_{3}+15r_{b})\right]
\label{eq84}%
\end{equation}
and
\begin{equation}
x_{-}=\tfrac{1}{5}\left[  6r_{1}^{3}+r_{1}^{2}(8r_{3}-25r_{b})-8r_{1}%
r_{3}(3r_{3}-5r_{b})-5r_{b}^{3}\right]  .\label{eq85}%
\end{equation}
Their comparison yields the relation
\begin{equation}
\left(  r_{1}+4r_{3}\right)  _{m=0}=5r_{b}\label{eq86}%
\end{equation}
which permits us to eliminate $r_{3}$ from Eq.~(\ref{eq83}) to obtain
\begin{equation}
t=\tfrac{1}{20}(r_{1}-r_{b})^{2}.\label{eq87}%
\end{equation}
Hence
\begin{equation}
\left.  r_{1}\right|  _{m=0}=r_{b}-2\sqrt{3t}\label{eq88}%
\end{equation}
and again with the use of Eq.~(\ref{eq77}) we obtain
\begin{equation}
\left.  r_{2}\right|  _{m=0}=\left.  r_{3}\right|  _{m=0}=r_{b}+\tfrac{1}%
{2}\sqrt{3t}.\label{eq89}%
\end{equation}
Substitution of these formulae into (\ref{eq84}) or (\ref{eq85}) yields the
motion law of the left boundary
\begin{equation}
x_{-}(t)=-6r_{b}t-12\sqrt{3}\,t^{3/2}.\label{eq90}%
\end{equation}
The plots of $x_{+}(t)$ and $x_{-}(t)$ are depicted in Fig.~4. To the right
from $x_{+}(t)$ and to the left from $x_{-}(t)$ the wave is described by the
dispersionless solution (\ref{eq59}). Between $x_{+}(t)$ and $x_{-}(t)$ we
have the region of fast oscillations represented by Eq.~(\ref{eq58}) with
$r_{i}(x,t)$, $i=1,2,3,$ given implicitly by Eqs.~(\ref{eq72}). The dependence
of $u$ on $x$ at some fixed moment of time is shown in Fig.~5. It
describes dissipationless shock wave connecting two smooth regions where we
can neglect dispersion effects. At the right boundary the periodic wave tends
to a sequence of separate dark soliton solutions of the mKdV equation and at
the left boundary the amplitude of oscillations tends to zero.

\begin{figure}[h]
\label{fig3} \centerline{
{\includegraphics{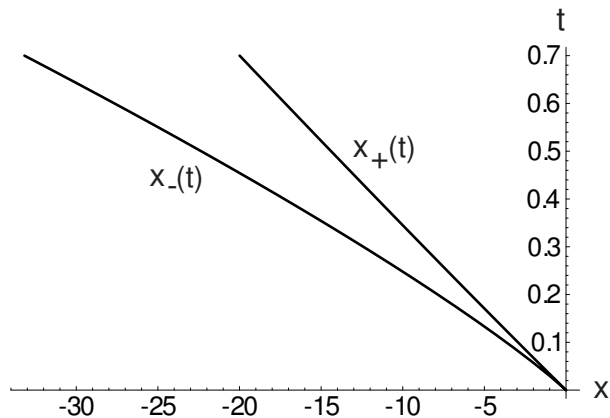}}} \caption{Dependence
of coordinates $x_{\pm} $ of boundaries of the region of oscillations on time
for the mKdV equation case.}%
\end{figure}

\begin{figure}[h]
\label{fig4} \centerline{
{\includegraphics{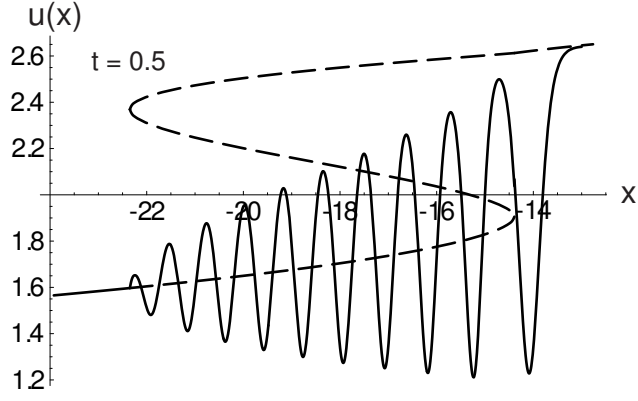}}} \caption{The
dissipationless shock wave for the mKdV equation. The parameters are equal to
$r_{b}=5$ and $t=0.5$. The dashed line represents the squared roots of the
Riemann invariants which match the smooth solution $u(x,t)$ of the
dispersionless equation at the boundaries of the region of oscillations.}%
\end{figure}

\subsection{Dissipationless shock wave in the KdV(2) equation (\ref{eq28})}

\label{shock_KdV2}

The theory of dissipationless shock wave for the KdV(2) equation is similar to
that for the KdV and mKdV cases. Therefore we shall present here only its main points.

The periodic solution of KdV(2) equation has the same form as in KdV equation
case (see, e.g. \cite{kamch2000}),
\begin{equation}
\label{eq91}u(x,t)=r_{2}+r_{3}-r_{1}-2(r_{2}-r_{1}){\mathrm{sn}} ^{2}\left(
\sqrt{r_{3}-r_{1}}\,\theta,m\right) , \quad m=\frac{r_{2}-r_{1}}{r_{3}-r_{1}},
\end{equation}
with phase velocity
\begin{equation}
\label{eq92}\theta=x-Vt,\quad V=2s_{2}-\tfrac32s_{1}^{2},
\end{equation}
corresponding to the second equation of the KdV hierarchy.

Now the periodic solution (\ref{eq91}) is parameterized by the Riemann
invariants $r_{i}$, $i=1,2,3,$ rather than by their squared roots, as it was
in mKdV equation case. Hence, the solution of the dispersionless equation
\begin{equation}
u_{t}=\tfrac{15}{2}u^{2}u_{x}\label{eq93}%
\end{equation}
near the wave breaking point should be taken in the form
\begin{equation}
x+\tfrac{15}{2}u^{2}=-(u-u_{b})^{3}.\label{eq94}%
\end{equation}
In fact, this form is equivalent near the wave breaking point to the solution
(\ref{eq56}), since $(u^{2}-u_{b}^{2})\simeq2u_{b}(u-u_{b})$ and the constant
factor can be scaled out. Formation of
multi-valued region is illustrated in Fig.~6. After taking into account the
dispersion effects it should be replaced by the dissipationless shock wave.

\begin{figure}[h]
\label{fig5} \centerline{
{\includegraphics{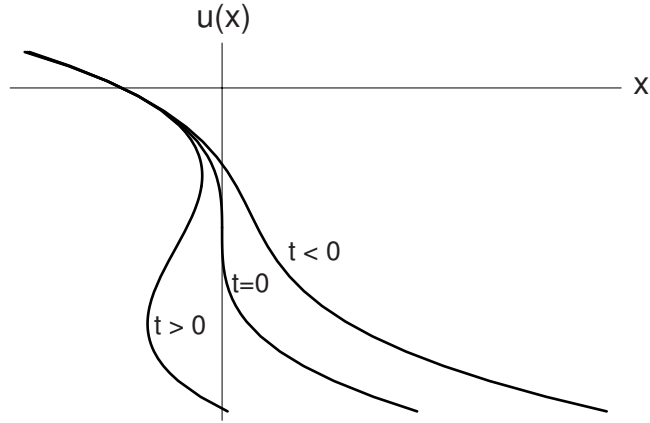}}} \caption{Formation of
multi-valued solution of the KdV(2) equation in dispersionless limit
(\ref{eq93}). The initial data correspond to the cubic curve $x=-(u-u_{b}%
)^{3}$ (see Eq.~(\ref{eq94})).}%
\end{figure}

\begin{figure}[h]
\label{fig6} \centerline{
{\includegraphics{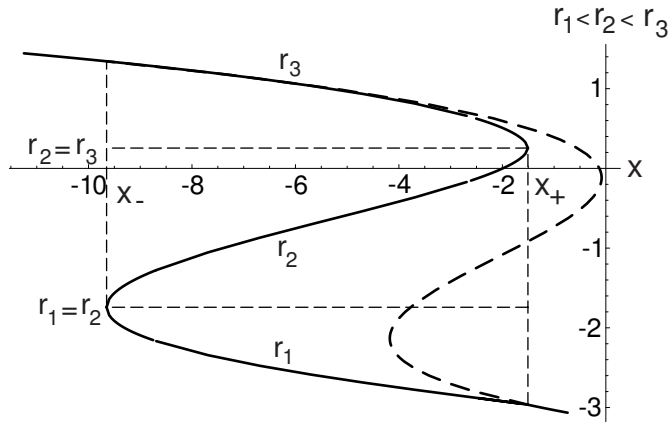}}} \caption{Dependence
of the Riemann invariants $r_{1},r_{2},r_{3}$ on $x$ at some fixed value of
time for the KdV(2) equation case. The plots are calculated according to
formulae (\ref{eq95}) with $u_{b}=-0.5$ and $t=0.25$. A dashed line represents
the dispersionless solution which matches the Riemann invariants at the
boundaries $x_{\pm} $ of the region of oscillations.}%
\end{figure}

\begin{figure}[h]
\label{fig7} \centerline{
{\includegraphics{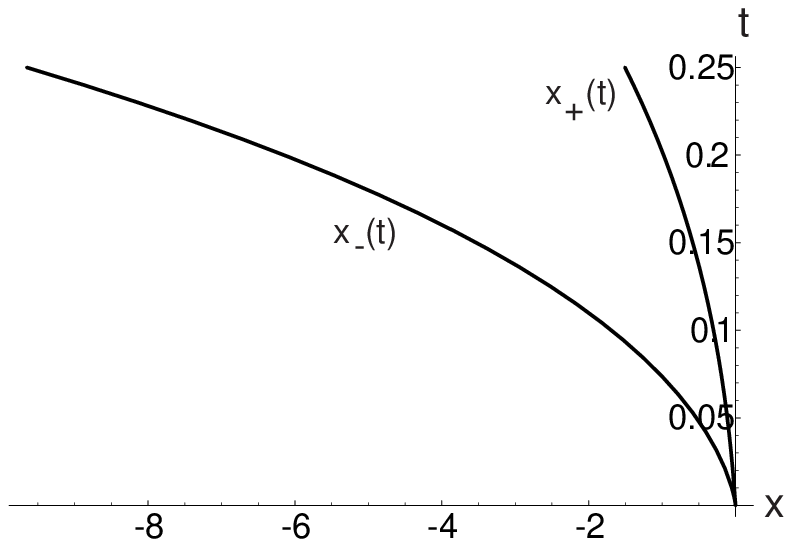}}} \caption{Dependence
of coordinates $x_{\pm} $ of boundaries of the region of oscillations on time
for the KdV(2) equation case.}%
\end{figure}

\begin{figure}[h]
\label{fig8} \centerline{
{\includegraphics{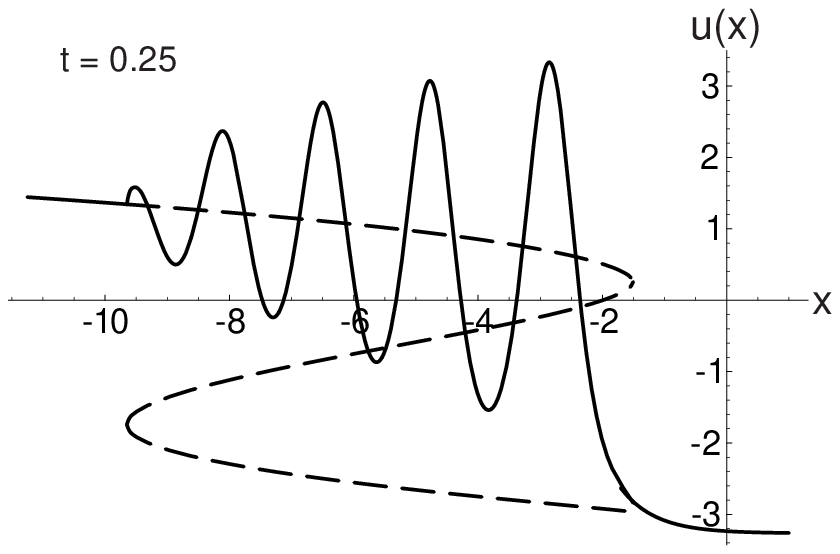}}} \caption{The
dissipationless shock wave for the KdV(2) equation. The parameters are equal
to $u_{b}=-0.5$ and $t=0.25$. The dashed line represents the Riemann
invariants which match the smooth solution $u(x,t)$ of the dispersionless
equation at the boundaries of the region of oscillations.}%
\end{figure}

Within the shock wave we have modulated periodic solution (\ref{eq91}) where
$r_{i}$ are slow functions of $x$ and $t$ and their evolution is governed by
the Whitham equations (\ref{eq52}) with $V$ defined by Eq.~(\ref{eq92}). Their
solution subject to the necessary boundary conditions can be found by the same
method as was used in the preceding subsection. As a result we obtain
\begin{equation}
x-w_{i}^{(2)}t=u_{b}^{3}+\tfrac{1}{2}u_{b}^{2}w_{i}^{(1)}-\tfrac{2}{5}%
u_{b}w_{i}^{(2)}+\tfrac{4}{35}w_{i}^{(3)},\label{eq95}%
\end{equation}
where $w_{i}^{(n)}$ are defined by formulae (\ref{eq66})--(\ref{eq68}).
Equations (\ref{eq95}) define implicitly the dependence of the Riemann
invariants $r_{i},$ $i=1,2,3,$ on $x$ and $t$. The resulting plots are shown
in Fig.~7. At the right boundary $x_{+}$ we have soliton limit ($m=1$) and at
the left boundary $x_{-}$ we have a wave with vanishing modulation.

The motion laws $x_{\pm}(t)$ can be found as in the preceding subsection, but
the final formulae become now quite complicated and we shall not write them
down. The corresponding plots are presented in Fig.~8. Again the region
between $x_{-}(t)$ and $x_{+}(t)$ corresponds to expanding with time $t$
dissipationless shock wave. It is illustrated in Fig.~9 where the dependence
$u(x)$ at fixed moment $t$ is shown. Now at one boundary bright solitons are
formed and at the other boundary amplitude of oscillations tends to zero.

\section{Conclusion}

We have shown that GDNLS equation with finite density boundary conditions can
be reduced, depending on values of the parameters $\eta$ and $\rho$, to
several important continuous models---KdV, mKdV, KdV(2) and 
fifth-order KdV equations
which describe different regimes of wave propagation in nonlinear lattice
(Salerno model). The KdV(2) equation appears at such values of the parameters
for which nonlinear and dispersive terms in in the KdV equation vanish, so
that the main contribution in small amplitude long wave approximation is given
by the third order nonlinear and fifth-order dispersion effects. This point
correspond to the AL equation case, and since the AL equation is completely
integrable and multi-scale method preserves complete integrability, we arrive
at the second equation of the KdV hierarchy. This observation explains
physical meaning of higher equations of integrable hierarchies---they give
main contribution into wave dynamics if lower order effects vanish in small
amplitude approximation of initial integrable equation.

The evolution equations obtained as approximations to the GDNLS equation lead
in dispersionless limit for general enough initial pulses to wave breaking so
that taking into account small dispersion effects leads to formation of
dissipationless shock waves. We have developed the theory of these waves in
framework of Whitham averaging method. Analytical expressions are obtained
which describe their main characteristics---trailing and leading end points,
amplitudes and wavelengths.

The phenomena described in the present paper are not restricted by the GDNL,
but are characteristic features of a large class of nonlinear Schr\"{o}dinger
lattices, which depend on one or more free parameters.

\acknowledgments 

The work of A.M.K. in Lisbon has been supported by the Senior NATO fellowship.
A.M.K. thanks also RFBR (grant 01-01-00696) for partial support. Work of A.S.
has been supported by the FCT fellowship SFRH/BPD/5569/2001. 
V.V.K. acknowledges
support from the European grant, COSYC n.o. HPRN-CT-2000-00158.

\end{document}